\documentclass{aastex631}

\usepackage{tablefootnote}
\usepackage{gensymb}

\shorttitle{Magnetic field Measurements of Stellar Coronae}
\shortauthors{Liu et al.}

\newcommand{\figa}[1]{Figure\,~\ref{#1}}

\newcommand{\eqn}[1]{Equation\,(\ref{#1})}

\definecolor{orange}{rgb}{1,0.4,0.}
\graphicspath{{./}{figures/}}

\begin{document}

\title{Forward Modeling of Magnetic-field Measurements at the Bases of Stellar Coronae through Extreme-Ultraviolet Spectroscopy}

\correspondingauthor{Hui Tian}
\email{huitian@pku.edu.cn}

\author{Xianyu Liu}
\affiliation{School of Earth and Space Sciences, Peking University, Beijing 100871, China}

\author{Hui Tian}
\affiliation{School of Earth and Space Sciences, Peking University, Beijing 100871, China}
\affiliation{State Key Laboratory of Space Weather, National Space Science Center, Chinese Academy of Sciences, Beijing 100190, China}
\affiliation{Key Laboratory of Solar Activity, National Astronomical Observatories, Chinese Academy of Sciences, Beijing 100012, China}

\author{Yajie Chen}
\affiliation{School of Earth and Space Sciences, Peking University, Beijing 100871, China}

\author{Wenxian Li}
\affiliation{Key Laboratory of Solar Activity, National Astronomical Observatories, Chinese Academy of Sciences, Beijing 100012, China}

\author{Meng Jin}
\affiliation{SETI Institute, 339 N Bernardo Ave suite 200, Mountain View, CA 94043, USA}

\author{Xianyong Bai}
\affiliation{Key Laboratory of Solar Activity, National Astronomical Observatories, Chinese Academy of Sciences, Beijing 100012, China}
\affiliation{School of Astronomy and Space Science, University of Chinese Academy of Sciences, Beijing 100049, China}

\author{Zihao Yang}
\affiliation{School of Earth and Space Sciences, Peking University, Beijing 100871, China}

\begin{abstract}

Measurements of the stellar coronal magnetic field are of great importance in understanding the stellar magnetic activity, yet the measurements have been extremely difficult.
Recent studies proposed a new method of magnetic field measurements based on the magnetic-field-induced-transition (MIT) of the Fe~{\sc{x}} ion.
Here we construct a series of stellar coronal magnetohydrodynamics (MHD) models and synthesize several Fe~{\sc{x}} emission lines at extreme-ultraviolet wavelengths, and then diagnose the magnetic field strength at the bases of the coronae using the MIT technique.
Our results show that the technique can be applied to some stars with magnetic fields more than three times higher than that of the Sun at solar maximum.
Furthermore, we investigate the uncertainty of the derived magnetic field strength caused by photon counting error and find that a signal-noise ratio of $\sim$50 for the Fe~{\sc{x}} 175 {\AA}~line is required to achieve effective measurements of the stellar coronal magnetic field.

\end{abstract}

\section{Introduction} \label{sec:intro}

The magnetic activity of host stars plays a pivotal role in determining the habitability of exoplanets \citep[e.g.,][]{Airpetian2017,Linsky2019}.
The stellar magnetic field is the energy source for a variety of explosive magnetic activity including superflares \citep[e.g.,][]{Maehara2012} and coronal mass ejections \citep[e.g.,][]{Argiroffi2019,Veronig2021,Namekata2022}.
These events could significantly affect the physical properties and chemical composition of exoplanetary atmospheres through various processes \citep[e.g.,][]{Cohen2014,Cherenkov2017,Tilley2019,Alvarado2022,Chen2021,Hazra2022}, which has put our understanding of the stellar magnetic field at the core of investigating the habitability of exoplanets.

Despite such importance, there have been only limited ways to measure the stellar magnetic field.
Most of the previous attempts only measured the stellar photospheric magnetic field based on Zeeman effect of spectral lines \citep[e.g.,][]{Johns2000,Donati2008,Morin2008,Reiners2012,Kochukhov2017,Kochukhov2019}.
Compared with the photospheric magnetic field, the coronal magnetic field is even more difficult to measure, which is the case for both the Sun and other stars \citep[e.g.,][]{Yang2020a,Yang2020b,Jiang2022,Zhu2022}.
There have been several attempts of stellar coronal magnetic field measurements based on radio observations \citep[e.g.,][]{Gary1981,Mutel1985,Gudel2002}, which could be subject to uncertainties because the radio emission mechanisms are often not easy to determine.
Indirect reconstructions of the stellar coronal magnetic field may be achieved through extrapolation from the photospheric magnetogram obtained with the Zeeman-Doppler imaging \citep[ZDI, see][]{Semel1989,Hebrard2016} technique.
However, this method can only reveal large-scale field structures, and uncertainties exist in both the extrapolation and ZDI techniques due to assumptions.

Recently, \citet{Li2015,Li2016} proposed that the intensity ratios of several Fe~{\sc{x}} lines at extreme-ultraviolet (EUV) wavelengths can be used to measure the solar coronal magnetic field strength based on the magnetic-induced-transition (MIT) theory.
\citet{Chen2021a} validated this technique for magnetic field measurements of solar active regions through forward modeling. Besides, this technique has been applied to actual spectral observations of solar active regions \citep[e.g.,][]{Si2020,Landi2020,Landi2021,Brooks2021}.
Inspired by these studies, our previous work \citep[][]{Chen2021b} suggested that the MIT technique can be extended to the measurements of the magnetic field strengths at the coronal bases of active stars in which the average photospheric magnetic flux density is at least one order of magnitude larger than that of the Sun.
However, in our previous study the input photospheric magnetograms of our models were scaled from a solar photospheric magnetogram taken during a relatively inactive phase of the solar cycle, which may not be appropriate for investigations of magnetic fields on active stars or solar-type stars at the peak of their long-term activity cycles \citep[e.g.,][]{Wilson1978,Bondar2019}.

In this paper, we start from a solar photospheric magnetogram taken during the solar maximum, and construct a series of stellar coronal models to explore the capability of the MIT technique in stellar coronal magnetic field measurements.
We also investigate what signal-to-noise ratio (S/N) is required to achieve effective measurements of the field strengths.
Section 2 describes the models used in this study, line synthesis, and methodology.
Our magnetic field diagnostic results are presented in section 3. In section 4 we investigate the uncertainties involved in this technique. Conclusions are given in section 5.

\section{Models and Emission Line Synthesis} \label{sec:models}

\subsection{Models} \label{subsec:models}

\begin{figure*} 
\centering {\includegraphics[width=\textwidth]{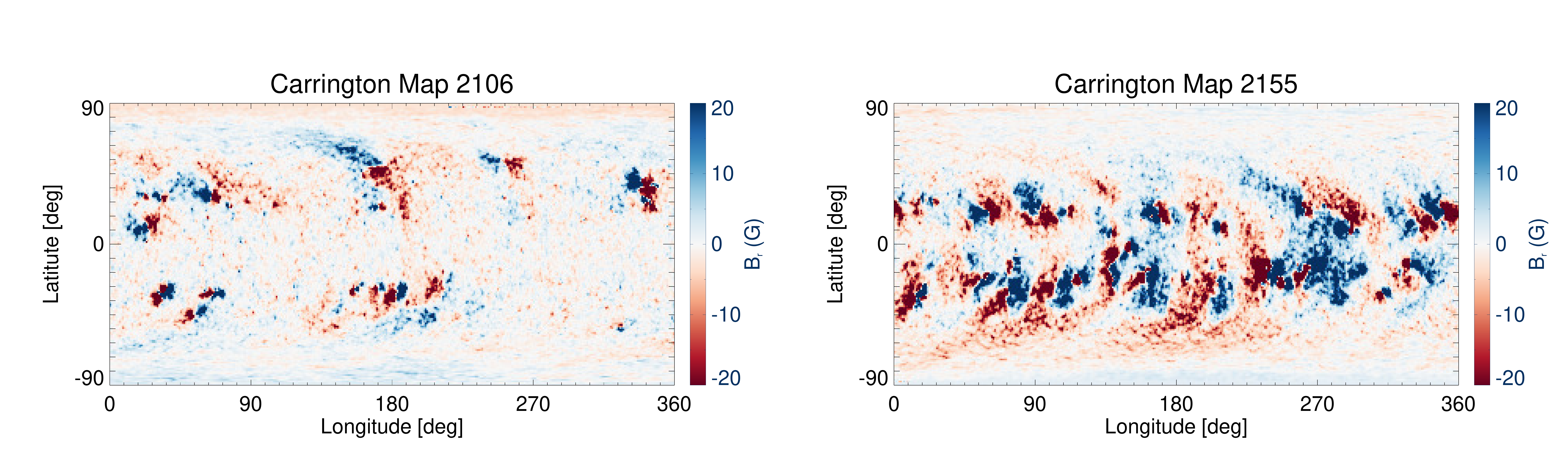}}
\caption{Photospheric radial magnetic field maps for CR 2106 and CR 2155. The field strength is saturated at $\pm$20 G. The total magnetic flux of CR 2155 map is $\sim5\times10^{23} $ Mx, which is $2-3$ times stronger than that of CR 2106 map ($\sim2\times10^{23} $ Mx).
} \label{fig1}
\end{figure*}

A series of stellar coronal models were constructed based on the Alfv{\'e}n Wave Solar Model \citep[AWSoM;][]{AWSoM} within the Space Weather Modeling Framework \citep[SWMF;][]{SWMF}. In this study, we only used the solar corona (SC) component since the emissions of the Fe~{\sc{x}} lines are mainly formed in the lower coronae or coronal bases of stellar atmospheres. This model assumes that the Alfv{\'e}n wave is injected at the inner boundary and the corona is heated through Alfv{\'e}n wave turbulence dissipation. The model also considers the radiative cooling effect and electron heat conduction. 

The inner boundary of the model was specified as the top of the chromosphere, where the electron temperature and number density were set to be $5\times10^4 $K and $2\times10^{17}$m$^{-3}$, respectively. The Poynting flux $S_{A}$ of the outgoing wave at the inner boundary is proportional to the radial magnetic field strength. This setting was made to satisfy the power-law relationship between the total coronal heating power and unsigned magnetic flux \citep[][]{Pevtsov2003,Sokolov2013}. For this reason, the radial magnetic field distribution at the inner boundary is required as input. As ZDI maps can reveal the large-scale magnetic field in the stellar photosphere, they have been chosen as input in some stellar simulations \citep[e.g.,][]{Cohen2014,Alvarado2018,Alvarado2019}. However, small-scale field structures are missed in ZDI maps, which would inevitably lead to an underestimation of the heating rate in the lower corona. To avoid this problem, in this work we used a synoptic solar magnetic field map obtained by the Global Oscillation Network Group (GONG) to resolve small-scale field structures. The magnetograms for stellar models are obtained through increasing the magnetic flux density by different factors.

In \citet{Chen2021b} we used synchronous field map on 2011 Feb 15, which corresponds to the Carrington Rotation 2106 (during the rising phase of solar cycle 24). In this study, we used the synoptic map of CR 2155 taken from 2014 Sep 17 to 2014 Oct 14 during the solar maximum. Both maps have the resolution of $1\degree$ at the disk center. The two magnetograms are shown in \figa{fig1} for comparison. We first constructed a solar CR 2155 model with the original CR 2155 magnetogram, then rescaled the CR 2155 magnetogram by different factors ranging from 2 to 9 and constructed a series of stellar coronal models. Each of the models is named ``xN", in which ``N" is the rescaling factor (``x1" represents the solar model). The transverse correlation length ($L_{\perp}$) of the Alfv{\'e}n waves is proportional to $1/\sqrt{B}$ \citep[][]{Hollweg1986}, which means $L_{\perp}\sqrt{B}$ is a constant in each model. In the ``x5", ``x6", ``x7", ``x8" and ``x9" models, $L_{\perp}\sqrt{B}$ was set to be the default value $1.5\times10^5 m\sqrt{T}$. For the models with weaker magnetic fields (``x1", ``x2", ``x3" and ``x4" models), to keep the models stable $L_{\perp}\sqrt{B}$ was adjusted to $3.0\times10^{5}m\sqrt{T}$. This ensures that the energy of Alfv{\'e}n wave is dissipated in a wider range of space, which can prevent the devastated accumulation of energy near the inner boundary and the chromospheric evaporation. The Alfv{\'e}n wave Poynting flux at the inner boundary is assumed to be proportional to the radial magnetic field, and the proportionality $(S_{A}/B)_{\odot}$ is another important parameter in the models. In the solar case, \citet{Sokolov2013} estimated $(S_{A}/B)_{\odot}$ to be $1.1\times10^{6}$ W/m$^{2}$. Here we assume that $(S_{A}/B)_{\odot}$ in the stellar models is the same as in the solar model. This leads to more energy injected into the coronae in the more active models. For more detailed descriptions of the numerical setups, we refer to \citet{Jin2012} and \citet{Oran2013}.

\begin{figure*} 
\centering {\includegraphics[width=0.9\textwidth]{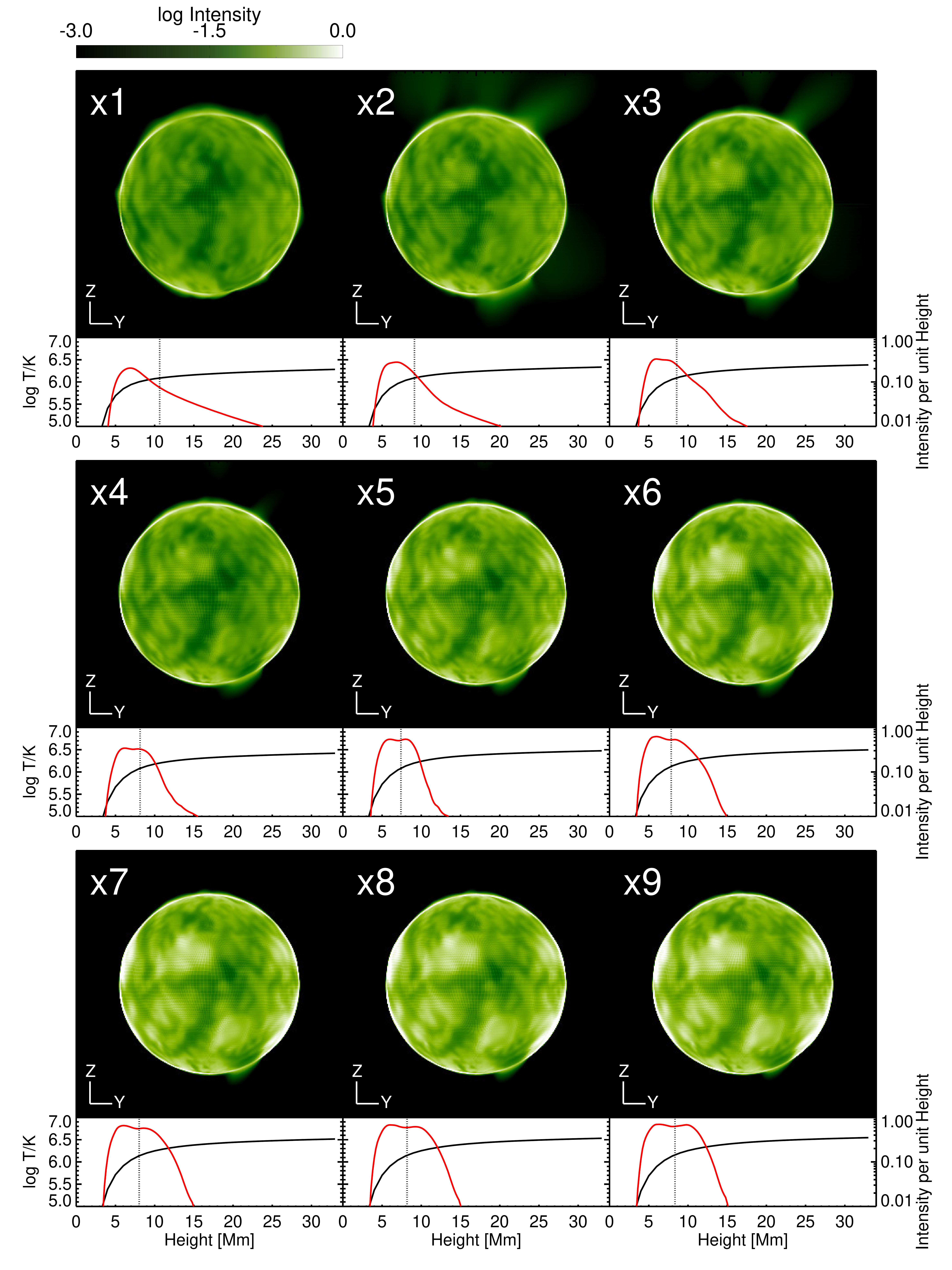}}
\caption{Fe~{\sc{x}} 257 {\AA} intensity images assuming that the LOS is along the $+x$-axis in different models. The intensity images are shown in logarithmic scale and arbitrary unit. The $+z$-axes are parallel to the stellar spin axis. The average temperature (black curve) and intensity (red curve) as a function of height above the stellar surface in each model are also shown. The vertical dashed line represents the emissivity-weighted height, as is defined in \eqn{eq1}.
} \label{fig2}
\end{figure*}

\subsection{Emission line synthesis} \label{subsec:emissions}

Similar to \citet{Chen2021a}, we first synthesized the emissions of the Fe x 174, 175, 177, 184, and 257 {\AA} lines from our models. The contribution functions were calculated using the CHIANTIPY code together with the atomic data from the CHIANTI database \citep[version 10.0;][]{Dere1997,DelZanna2021}, in which the collision data originate from \citet{DelZanna2012}. For the radiative transition data, we used the results obtained from the multiconfiguration Dirac-Hartree-Fock and relativistic configuration interaction calculations by \citet{Wang2020} and \citet{Li2021}. The MIT transition rates as a function of magnetic field strengths were obtained using the values given by \citet{Li2021}. We first created the contribution functions for the related Fe~{\sc{x}} lines at a large number of electron temperatures ($\log T/K$, from 5.6 to 6.6 with an interval of 0.004), electron densities ($\log N/cm^{-3}$, from 6 to 15 with an interval of 0.04), and magnetic field strengths ($B/Gauss$, from 0 to 6500 with an interval of 10 Gauss). We then interpolated the contribution functions to the $\log T/$K, $\log N/$cm$^{-3}$, and $B/$G values of the grid points in the models, and synthesized the spectral line emissions from different locations on the stars. \figa{fig2} presents the Fe~{\sc{x}} 174 {\AA} images synthesized from different models by assuming that the line of sight (LOS) is along the $+x$-axis. We also calculated the average temperature and intensity as a function of height in each model. 

The most significant difference among the models is the different coronal temperatures resulting from different heating rates. In the x1 (solar) model, the temperature reaches $10^{6.2-6.3} $ K at a height of $30$ Mm. In those active stellar models, the more intense heating at the inner boundary leads to higher coronal temperatures (e.g., $10^{6.5} $ MK in the x9 model). Because of this, the emissions at higher heights in the models with stronger magnetic fields are considerably weakened, due to the fact that the contribution functions of the Fe~{\sc{x}} lines peak at a temperature of $\sim1$ MK. We defined an average formation height of the Fe~{\sc{x}} 174 {\AA} emission line as the emissivity-weighted height:
\begin{equation}
    H_{0}=\frac{\int_{V_0} H \cdot e_{174} dV}{\int_{V_0} e_{174} dV} \label{eq1}
\end{equation}
where $e_{174}$ is the emissivity of the Fe~{\sc{x}} 174 {\AA} line and $V_0$ represents the whole simulation box. The average formation height of the 174 {\AA} line in each model is shown as a vertical dashed line in \figa{fig2}. In all the models, the average formation height is below $11$ Mm, which indicates that the integrated intensities of Fe~{\sc{x}} emissions mainly come from the low coronae or coronal bases. 

To mimic different viewing angles, we selected a series of LOS directions every $30\degree$ in both the inclinational and azimuthal directions. Since the stars are spatially unresolved, the emissivity of a spectral line was integrated over the whole domain except the far side of the stellar disk. In this way, we obtained the intensities of different Fe~{\sc{x}} lines and line ratios for each LOS. As predicted by the MIT theory, the intensity ratio between the Fe~{\sc{x}} 257 {\AA} line and another Fe~{\sc{x}} reference line (e.g., the 174, 177 or 184 {\AA} line) is a function of electron temperature $T$, density $N$ and magnetic field strength $B$. In this study, we chose a fixed temperature of $10^{6.02} $ K, at which the contribution functions of the Fe~{\sc{x}} lines peak \citep[][]{Li2021}. The electron density was derived from the intensity ratio of the density-sensitive 175/174 {\AA} line pair \citep[][]{Brosius1998,DelZanna2018}. After these, the magnetic field strength $B_1$ was derived from the intensity ratios of the Fe~{\sc{x}} lines. Another MIT diagnosing method called ``weak field strength technique" was introduced and used by \citet{Landi2020} to measure the relatively weak magnetic field strength (i.e., below $\sim150-200$ Gauss) in the solar corona. However, since most of the measured magnetic field strengths of our models exceed the upper limit of the measurable range (as one will see in \figa{fig3}), this method is not considered in this study.

To compare the measured values with the values in models, an average magnetic field strength in the model is needed. Since the emission lines only carry the information on the magnetic field in source regions of Fe~{\sc{x}} emissions, the average magnetic field strength in the simulation domain should be weighted by the Fe~{\sc{x}} emissivity. Therefore, we use the Fe~{\sc{x}} 174 {\AA} line emissivity-weighted average field strength, i.e.
\begin{equation}
    B_{0}=\frac{\int_{V(i)} B \cdot e_{174} dV}{\int_{V(i)} e_{174} dV} \label{eq2}
\end{equation}
to represent the average magnetic field strength in the source regions. Here $i$ is the $i_{th}$ LOS direction, and $V(i)$ is the whole simulation domain excluding the far-side region for each LOS.

\section{Magnetic field diagnostic result} \label{sec:results}



\begin{figure*} 
\centering {\includegraphics[width=0.9\textwidth]{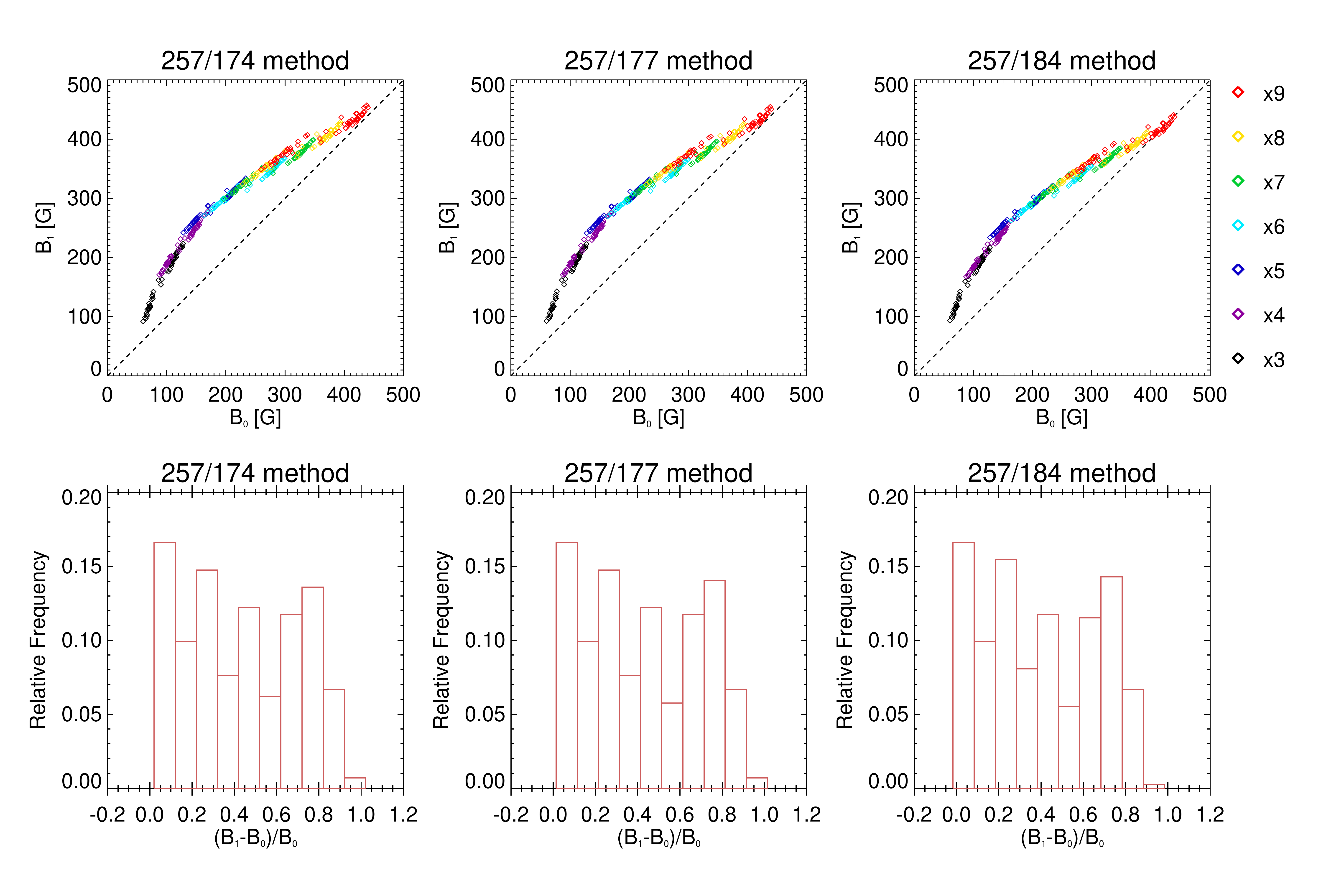}}
\caption{Measured field strength versus emissivity-weighted field strength in the models. Upper row: scatter plots for the relationship between $B_0$ and $B_1$. Each panel represents results derived from a line ratio. Different colors represent results from different models. The different data points of each color represent results obtained with different LOS directions. Lower row: histograms of $(B_{1}-B_{0})/B_{0}$.
} \label{fig3}
\end{figure*}

\figa{fig3} shows a comparison of the magnetic field strengths calculated using intensity ratios of Fe~{\sc{x}} 257/174 {\AA}, 257/177 {\AA}, and 257/184 {\AA} line pairs ($B_1$) and values in the model ($B_0$). In each panel of the upper row, we present the results derived from the same line pair for all the models. The black dashed line in each panel indicates $B_1=B_0$, i.e., the magnetic field strengths derived from line ratios are equal to values in the models. The ranges of magnetic field strength in the models x1--x9 ($B_0$) are $16-30$, $34-71$,  $60-128$,  $86-169$,  $127-234$,  $163-300$,  $196-348$,  $228-394$, and $259-441$ Gauss, respectively. As a representation of average magnetic strength in the source regions of the Fe~{\sc{x}} emissions, $B_0$ not only depends on the rescaling factor, but also depends on the distance from the stellar disk to the source regions, which is revealed by the formation height of Fe~{\sc{x}}. Thus, $B_0$ is not proportional to the rescaling factor since the formation height of the Fe~{\sc{x}} lines differs in different models. The density values ($\log N/cm^{-3}$) along different LOS directions of the models, as calculated using the F~{\sc{x}} 175/174 {\AA} line ratio, are  $8.6-8.73$,  $8.74-8.91$,  $8.88-9.06$,  $8.94-9.11$,  $9.13-9.27$,  $9.21-9.34$,  $9.25-9.37$,  $9.28-9.4$,  and $9.3-9.41$, respectively. For models with a surface magnetic flux density more than three times larger than the solar case (x3--x9 models), the MIT method was found to be capable of yielding reasonable estimations of the coronal magnetic field strengths for all the LOS directions. However, this technique fails to provide reasonable estimations for almost all LOS directions in the x1 and x2 models, since the obtained intensity ratios of the Fe~{\sc{x}} 257 {\AA} and the reference lines do not fall in the range predicted by the MIT theory, a phenomenon that is most likely due to the low densities ($N<10^{8.8} $cm$^{-3}$) in these two models (see a more detailed discussion in \citet{Chen2021b}.

\begin{figure*} 
\centering {\includegraphics[scale=0.5]{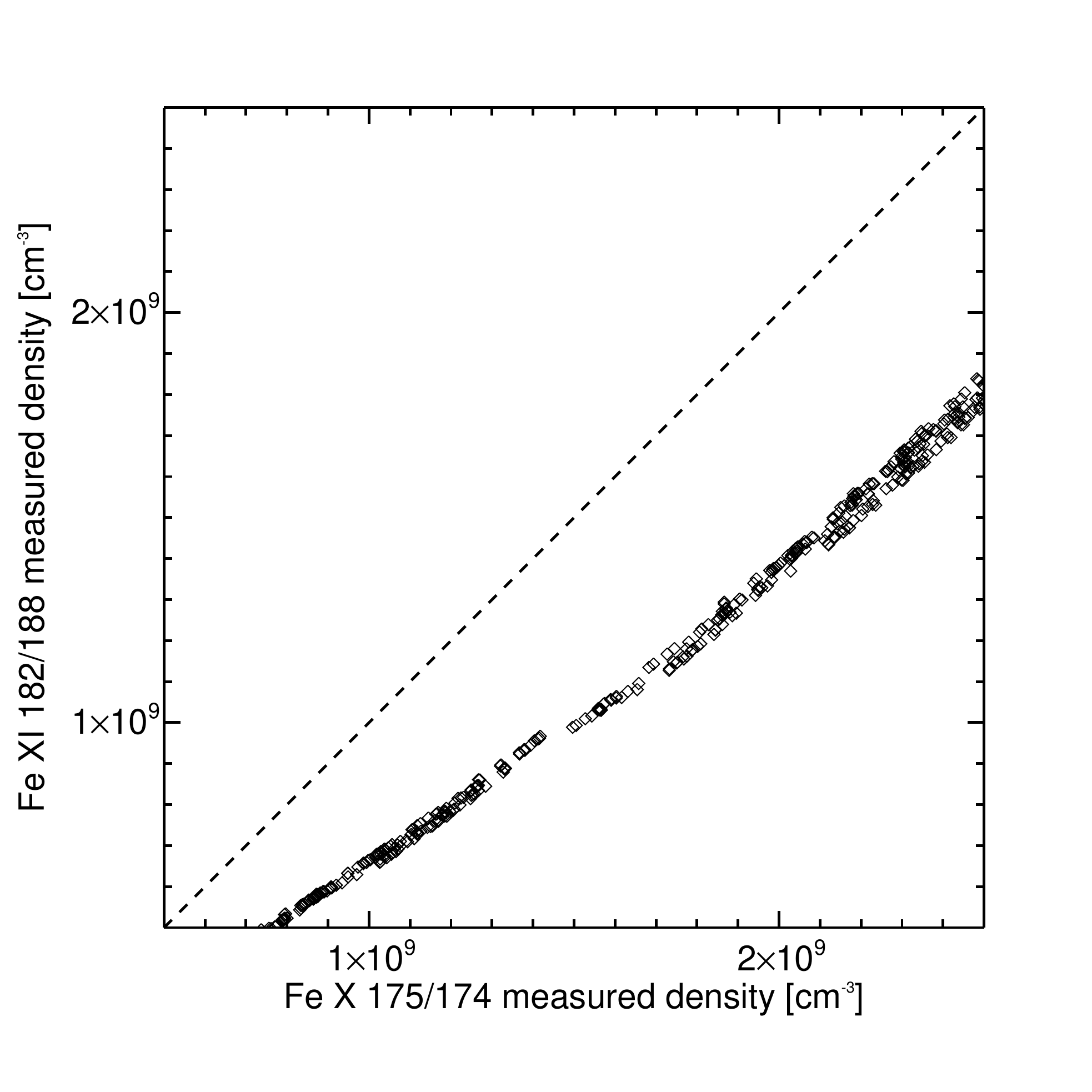}}
\caption{
Comparison between the densities measured by the Fe~{\sc{x}} 175/174 line pair and by the Fe~{\sc{xi}} 182.17/(188.22+188.30) {\AA} intensity ratio.
} \label{fig4}
\end{figure*}

\begin{figure*} 
\centering {\includegraphics[width=\textwidth]{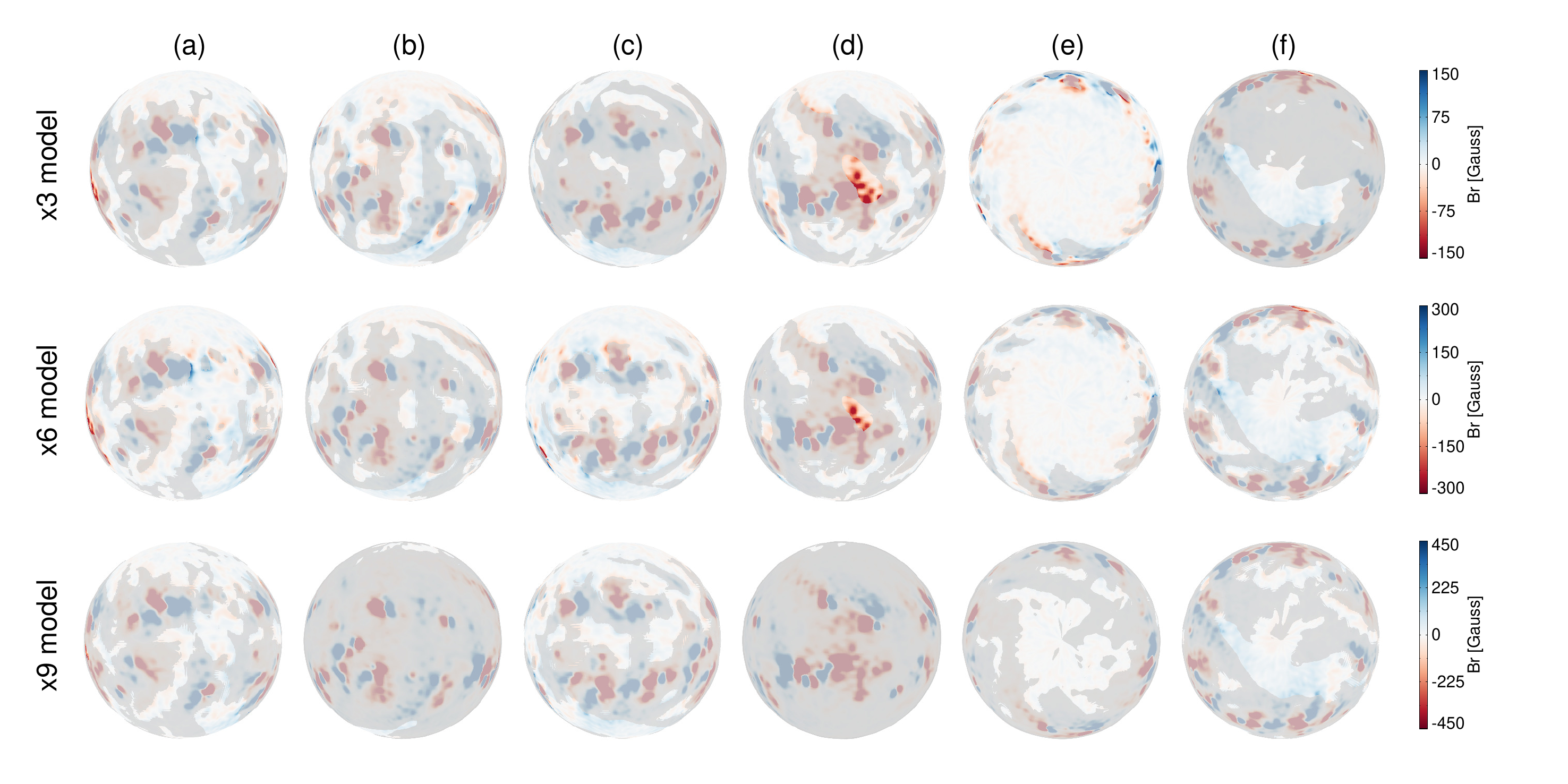}}
\caption{Best fitted $V^{'}$ (grey isosurface) for six LOS directions in the x3, x6 and x9 models. From (a) to (f), the LOS directions are along $\pm x, \pm y, \pm z$-axes, respectively. The radial magnetic field strength is saturated at $\pm150$ G, $\pm300$ G, and $\pm450$ G for the three models, respectively.
} \label{fig5}
\end{figure*}

As one will see in section 4, the Fe~{\sc{x}} 175 {\AA} line used for density measurements is relatively weaker compared to other Fe~{\sc{x}} lines, which means that in actual observations the S/N ratio of the Fe~{\sc{x}} 175 {\AA} line will be smaller. \citet[][]{Landi2020} used the Fe~{\sc{xi}} 182.17/(188.22+188.30) {\AA} intensity ratio as a substitute for the Fe~{\sc{x}} 175/174 {\AA} line ratio to measure the density. To explore the suitability of this method for our stellar investigations, we calculated the contribution functions of Fe~{\sc{xi}} 182.17, 188.22 and 188.30 {\AA} lines, and synthesized their emissions in the model. We then used the Fe~{\sc{xi}} intensity ratio to measure the electron density. \figa{fig4} shows the comparison between the densities measured by the Fe~{\sc{x}} and Fe~{\sc{xi}} intensity ratios. The density measured by the Fe~{\sc{xi}} intensity ratio is only $69\%$ of the density from the Fe~{\sc{x}} line pair on average. This is because the contribution functions of the Fe~{\sc{xi}} lines peak at a higher temperature ($10^{6.12} $ K) compared to those of Fe~{\sc{x}} lines, which means that the Fe~{\sc{xi}} lines are formed at higher heights where the density is smaller. We also found that the derived magnetic field strength using the density from the Fe~{\sc{xi}} intensity ratio is underestimated by $\sim200$ Gauss compared with $B_0$, and the correlation between $B_1$ and $B_0$ is weaker compared to that using the density from the Fe~{\sc{x}} 175/174 {\AA} line pair. Therefore, the magnetic field measurement result using the density from the Fe~{\sc{xi}} intensity ratio is not shown here. Below we still use the Fe~{\sc{x}} 175/174 {\AA} line pair to measure the density.

From \figa{fig3}, one can clearly see that $B_1$ monotonically increases with $B_0$ in general. However, one may also notice that $B_1$ deviates from $B_0$ for almost all the cases. The lower row of \figa{fig3} shows the frequency histograms of the relative error $(B_1-B_0)/B_0$. In each histogram, $(B_1-B_0)/B_0$ is around 0.4 on average. It is worth mentioning that $B_0$ is empirically defined, thus $B_0$ and $B_1$ are not theoretically identical. For this reason, the plots of $B_1$ versus $B_0$ only serve to reveal the general relevance of the measured field strengths with the real values. In order to investigate the actual physical meaning of the measured field strength $B_1$, we attempted to find out the region in which the emissivity-weighted average field strength is equal to $B_1$. To do this, we chose different emissivity thresholds $e_{th}$ and delimited the corresponding spaces $V^{'}$ in which the emissivity is higher than $e_{th}$. For the $i_{th}$ LOS and an emissivity threshold of $e_{th}$, the emissivity-weighted average field strength within $V^{'}$ is then defined as
\begin{equation}
    B_0(i,e_{th})=\frac{\int_{V^{'}(i,e_{th})} B \cdot e_{174} dV}{\int_{V^{'}(i,e_{th})} e_{174} dV} \label{eq3}
\end{equation}
The difference between \eqn{eq2} and \eqn{eq3} is that \eqn{eq2} calculates the average magnetic field in the whole space, while \eqn{eq3} calculates the average magnetic field only in regions with strong Fe~{\sc{x}} emissivity. One can derive $e_{th}$ by solving $B_0(i,e_{th})=B_1(i)$. In \figa{fig5}, we present $V^{'}$ as grey isosurfaces above the photosphere for six different LOS directions in the x3, x6, and x9 models. Note that the radial extension of each isosurface is very small and the height of the isosurface varies with location, which cannot be shown in these two-dimensional images. Nevertheless, we can still see that in most of the cases the isosurface of $V^{'}$ significantly overlaps with the active regions. Thus, it appears that the measured field strength largely reflects the average field strength in stellar active regions.

From the analyses above, we may conclude that the MIT method can be applied to stars with a field strength of at least 3 times larger than that of the Sun during solar maximum (i.e., stars with $B_0$ of at least $\sim60$ Gauss). In \citet{Chen2021b}, we suggested that the MIT method could be applied to stars with a field strength of at least 20 times larger than that of the Sun (i.e., stars with $B_0$ of at least $\sim160$ Gauss). This discrepancy is mainly due to the difference of densities in models used by the two studies. \citet{Chen2021b} found that the MIT measurability of the magnetic field strength significantly increases with the stellar coronal density. As a comparison, the x3 model in this study and the x10 model in the previous study both have an average $B_0$ of $\sim80-90$ Gauss, while the average measured coronal density of the former model ($10^{9.04} $cm$^{-3}$) is more than $1.7$ times larger than that of the latter model ($10^{8.80} $cm$^{-3}$). This indicates that the stellar models in this study are generally denser in the corona as compared to the models in the previous study under the same magnetic field strength. The reason for such a difference is that compared to the CR2106 magnetogram used by the previous study, the CR2155 magnetogram in this study has more active regions, in which the density is higher than that in the quiet region. Therefore, the higher stellar densities can facilitate measurements of magnetic field strength when the stellar activity level is high.

\section{Effect of Signal-to-Noise Ratios on Magnetic field measurements} \label{noise}

\begin{figure*} 
\centering {\includegraphics[width=0.9\textwidth]{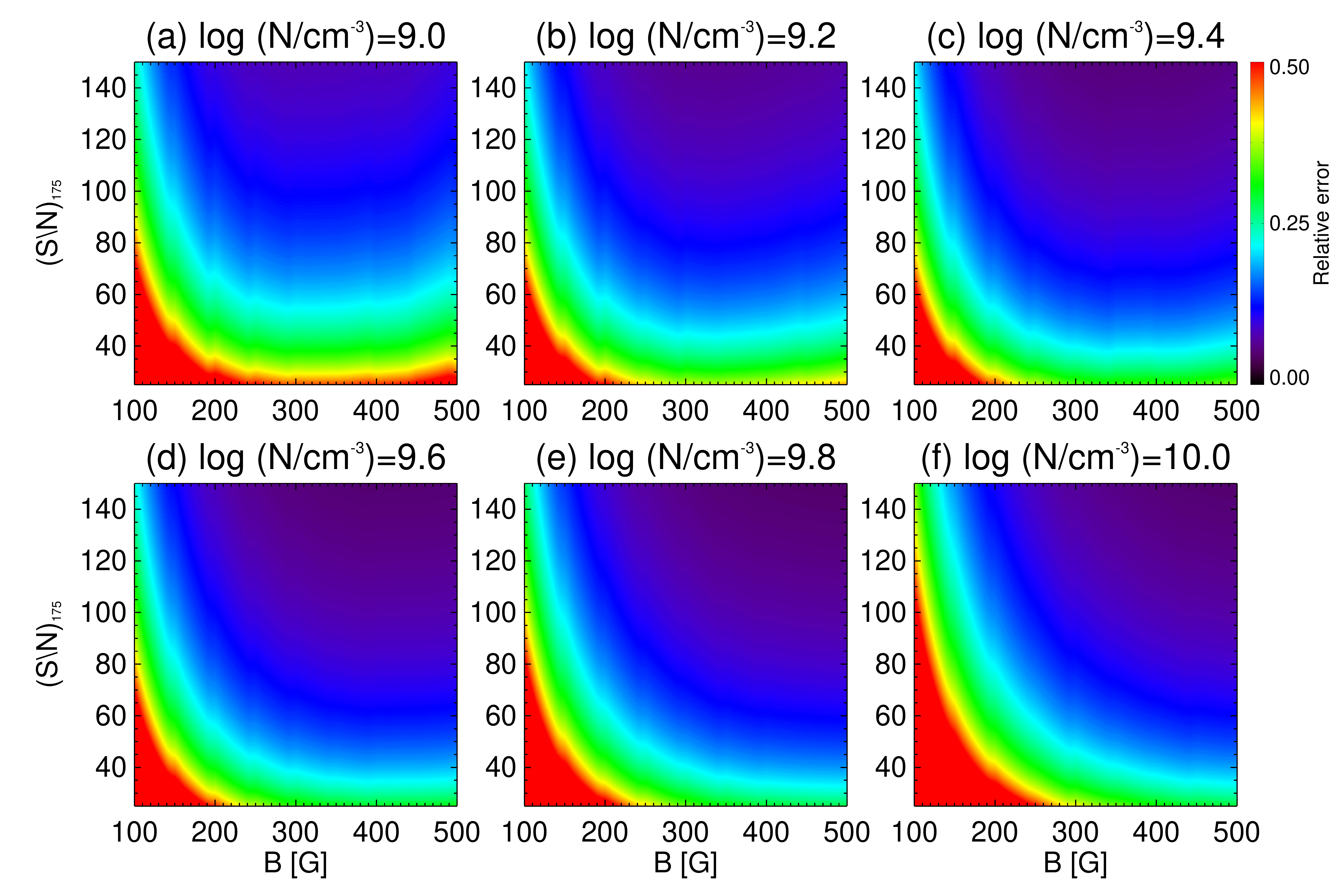}}
\caption{Relative error of measured magnetic field strength as a function of density, magnetic field strength and $(S/N)_{175}$. Each panel corresponds to a specific density.
} \label{fig6}
\end{figure*}

\begin{figure*} 
\centering {\includegraphics[width=\textwidth]{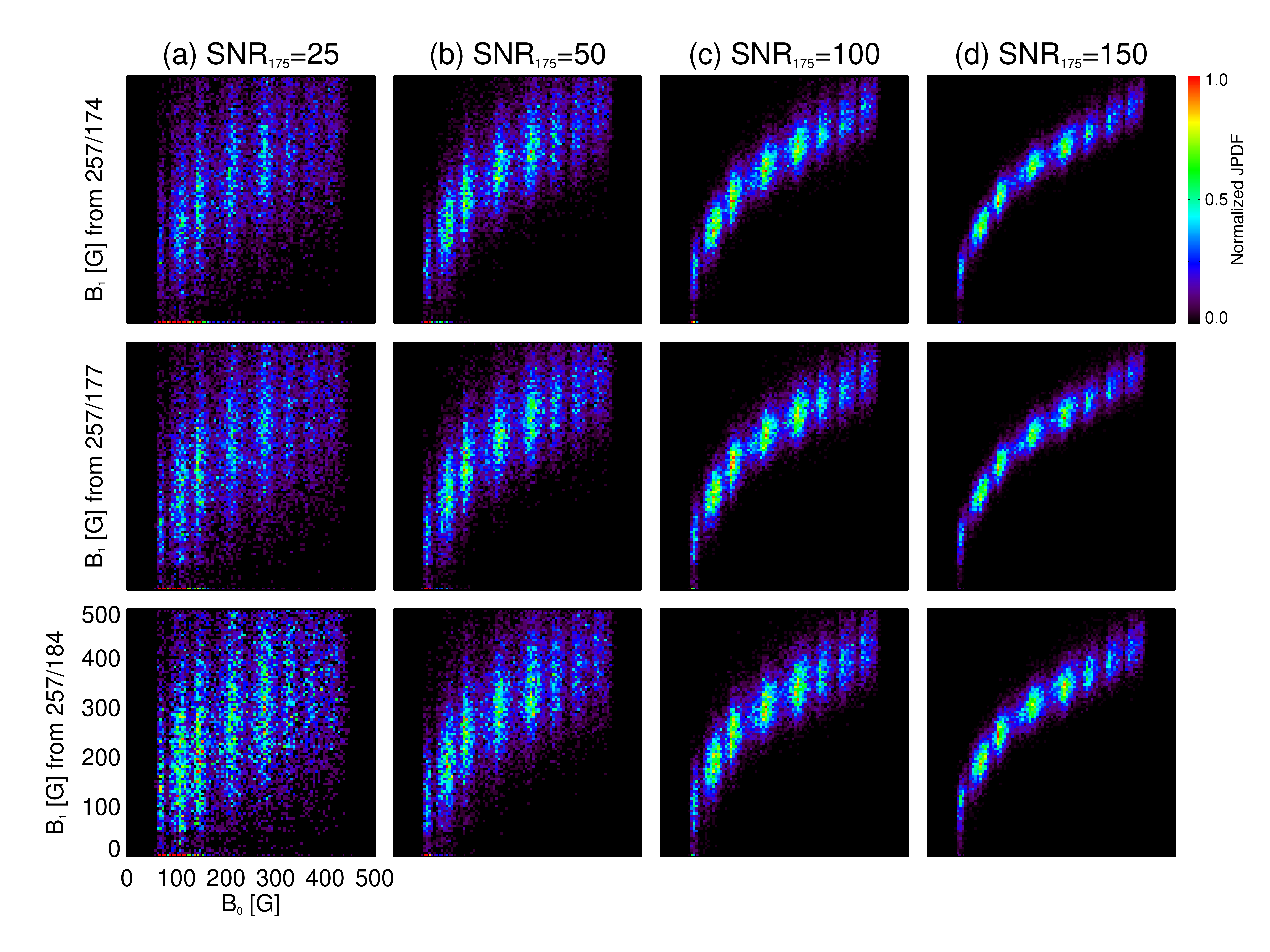}}
\caption{JPDF of measured magnetic field strength versus magnetic field strength in the model. Each column represents results under a specific noise level.
} \label{fig7}
\end{figure*}

So far, we have not yet considered the unavoidable uncertainties in real observations, which may affect the application of the MIT technique for the magnetic field measurements. In real observations, uncertainties could come from photon counting, calibration, and accuracy of atomic database. Here, we focus on the effect of noise caused by photon counting.

We first attempted to find out the uncertainty of the derived magnetic field strengths as a function of signal-noise ratio, density, and magnetic field strength. For simplification, we assumed that the source region of Fe~{\sc{x}} emissions has uniform physical parameters (i.e., magnetic field strength $B$, density $N$, and temperature $T$. $T$ is assumed to be $10^{6.02} $ K). The MIT technique uses the intensity ratio of the Fe~{\sc{x}} 175/174 {\AA} line pair ($r_1$) to calculate the density $N$, and uses the intensity ratio of the Fe~{\sc{x}} 257/$x$ ($x$ is an Fe~{\sc{x}} reference line at 174 {\AA}, 177 {\AA} or 184 {\AA}) line pair ($r_2$) to calculate the magnetic field strength $B$, i.e.
\begin{equation}
    B=B(r_2, N), N=(r_1) \label{eq4}
\end{equation}
The temperature is not explicit in the functions above since it is fixed to be $10^{6.02} $ K. The measured density and magnetic field strength can thus be estimated by their corresponding values in the source region. Differentiating \eqn{eq4} gives
\begin{equation}
    dB=(\frac{\partial B}{\partial r_2})_{N}dr_2+(\frac{\partial B}{\partial N})_{r_2}dN \label{eq5}
\end{equation}
\begin{equation}
    dN=\frac{dN}{dr_1}dr_1 \label{eq6}
\end{equation}
Here $dr_1$ and $dr_2$ represent the differentials of line ratios $r_1$ and $r_2$, respectively. $dN$ and $dB$ are the corresponding differentials of $N$ and $B$, respectively. $dr_1$ and $dr_2$ are expressed by differentials of line intensities as
\begin{equation}
    dr_1=d(\frac{I_{175}}{I_{174}})=r_1(\frac{dI_{175}}{I_{175}}-\frac{dI_{174}}{I_{174}}) \label{eq7}
\end{equation}
\begin{equation}
    dr_2=d(\frac{I_{257}}{I_{x}})=r_2(\frac{dI_{257}}{I_{257}}-\frac{dI_{x}}{I_{x}}) \label{eq8}
\end{equation}
Combining \eqn{eq5}-\eqn{eq8} we have
\begin{equation}
    \frac{dB}{B}=a(\frac{dI_{175}}{I_{175}}-\frac{dI_{174}}{I_{174}})+b(\frac{dI_{257}}{I_{257}}-\frac{dI_{x}}{I_{x}}) \label{eq9}
\end{equation}
where $a=\frac{r_1}{B}(\frac{\partial B}{\partial N})_{r_2}/\frac{dr_1}{dN}$, $b=\frac{r_2}{B}(\frac{\partial B}{\partial r_2})_{N}$. Since $B$ is a binary function of $N$ and $r_2$, we can rewrite $(\frac{\partial B}{\partial N})_{r_2}$ and $(\frac{\partial B}{\partial r_2})_{N}$ as $-(\frac{\partial r_2}{\partial N})_{B}/(\frac{\partial r_2}{\partial B})_{N}$ and $1/(\frac{\partial r_2}{\partial B})_{N}$, respectively. Here $(\frac{\partial r_2}{\partial B})_{N}$, $(\frac{\partial r_2}{\partial N})_{B}$, and $\frac{dr_1}{dN}$ can be calculated from the intensity ratios of Fe X lines as a function of magnetic field strength and density, thus both $a$ and $b$ are derived. We assume that the intensity of each line is subject to an uncertainty (e.g., the relative uncertainty of $I_{174}$ is $\delta_{174}$) caused by photon counting. The corresponding relative error of magnetic field strength is given by
\begin{equation}
    \delta_B=\sqrt{a^{2}(\delta_{174}^2+\delta_{175}^2)+b^{2}(\delta_{x}^2+\delta_{257}^2)}\label{eq10}
\end{equation}
The relative uncertainty of each line equals the reciprocal of S/N. As the 175 {\AA} line is the weakest line among all the Fe~{\sc{x}} lines we used in this study, the S/N of the 175 {\AA} line ($(S/N)_{175}$) was used as a representation of noise level of actual observations. Since the S/N equals the square root of photon number n, the S/N of another line y (y can be 174, 177, 184, or 257 {\AA} line) is
\begin{equation}
    (S/N)_{y}=(S/N)_{175}\cdot \sqrt{\frac{n_y}{n_{175}}}\label{eq11}
\end{equation}
After these, $\delta_B$ is then rewritten as
\begin{equation}
    \delta_B=\frac{1}{(S/N)_{175}}\cdot \sqrt{a^{2}(\frac{n_{175}}{n_{174}}+1)+b^{2}(\frac{n_{175}}{n_x}+\frac{n_{175}}{n_{257}})}\label{eq12}
\end{equation}
Here $a$, $b$ and all the photon number ratios are functions of $B$ and $N$, thus $\delta_B$ is a function of $(S/N)_{175}$, $B$, and $N$.

\begin{table*}
\caption{The $S/N$ of 174, 175, 177, 184, 257 {\AA} lines when assuming $(S/N)_{175}=50$ and the LOS direction is along the $+x$-axis in ``x3''- ``x9'' models.}\label{tab1}
\begin{center}
\setlength{\tabcolsep}{0.72cm}
\begin{tabular}{c|ccccccc}
\hline\hline
Model & x3 & x4 & x5 & x6 & x7 & x8 & x9\\
\hline
$(S/N)_{174}$ & $156.9$ & $150.0$ & $132.7$ & $125.9$ & $122.7$ & $120.3$ & $118.4$ \\
$(S/N)_{175}$ & $50.0$ & $50.0$ & $50.0$ & $50.0$ & $50.0$ & $50.0$ & $50.0$ \\
$(S/N)_{177}$ & $119.1$ & $114.9$ & $100.8$ & $95.7$ & $93.3$ & $91.5$ & $90.1$ \\
$(S/N)_{184}$ & $76.1$ & $73.0$ & $65.2$ & $62.1$ & $60.7$ & $59.6$ & $58.8$ \\
$(S/N)_{257}$ & $81.6$ & $77.6$ & $66.9$ & $62.8$ & $61.0$ & $59.7$ & $58.7$ \\
\hline
\end{tabular}
\end{center}
\end{table*}

In this section we chose the 174 {\AA} line as the reference line when diagnosing the magnetic field strength. We chose six different values of $\log(N/cm^{-3})$ ranging from $9.0$ to $10.0$, and plot $\delta_B$ as a function of $(S/N)_{175}$ and $B$ under each specified density value in \figa{fig6}. The most obvious feature is that $\delta_B$ decreases with magnetic field strength, which would facilitate observations of stars with stronger fields. Meanwhile, the $\delta_B$ does not change significantly with density. For stars with weaker magnetic fields, the derived magnetic field strengths are expected to be subject to larger uncertainties.

To investigate what S/N is required when applying the MIT technique, we added different levels of noise above the emissions of the Fe~{\sc{x}} lines. The intensity of a spectral line was then modified to
\begin{equation}
    I_x^{'}=I_x\cdot (1+\frac{R}{S/N_{175}})\label{eq13}
\end{equation}
where $I_x$ is line intensity primarily calculated in section 3, and $I_{x}^{'}$ is the modified intensity after considering the noise. The subscript $x$ ($x$=174,175...) represents the wavelength of the Fe~{\sc{x}} line in the unit of {\AA}. $R$ is a random value given by Gaussian distribution that has a standard deviation of $1$ and an average value of $0$. We chose a series of $(S/N)_{175}$ as $25$, $50$, $100$, and $150$, and in each case the S/N values of other lines were determined by $(S/N)_{175}$ and photon number ratios. The S/N values of all the Fe~{\sc{x}} lines when assuming that $(S/N)_{175}=50$ and the LOS direction is along $+x$-axis are shown in Table 1 as an example. The diagnosing procedure described in section 3 was then repeated for all the x3--x9 models. To increase the number of data points, the LOS directions were taken every $5\degree$ for both the inclination and azimuth. \figa{fig7} shows the results in joint probability density functions of $B_1$ and $B_0$. We also calculated the Spearman correlation coefficient $\rho$ between $B_1$ and $B_0$ when choosing the 174 {\AA} line as the reference line. We found that $\rho$ is $0.40$, $0.73$, $0.89$, and $0.93$ when $(S/N)_{175}$ is $25$, $50$, $100$, and $150$, respectively. We can see that a $(S/N)_{175}$ of $25$ appears to be insufficient for effective measurements, while a $(S/N)_{175}$ of $50$ or higher is enough to maintain a good correlation between the measured and real field strengths, as it is also suggested by \figa{fig6} that $\delta_B$ is around $20-30\%$ when $(S/N)_{175}$ is around $50$. Based on these analyses, we propose that a $(S/N)_{175}$ of at least $50$ should be guaranteed in real observation to achieve effective measurements of the coronal magnetic field.

\section{Summary and Conclusions} \label{conclusions}
In this work, we have investigated the potential of the MIT technique to measure the stellar coronal magnetic field strength and discussed the associated uncertainty caused by photon counting error. A series of stellar coronal models were constructed based on a solar photospheric synoptic magnetogram at solar maximum. By comparing the measured field strengths with values in the models, we concluded that the MIT technique can be used to measure the magnetic field strengths at the coronal bases of some stars with average photospheric field strengths of at least 3 times larger than that of the Sun at solar maximum. 

We have also derived an analytical expression for the dependence of the relative error of the measured field strength on the signal-noise ratio. By considering the photon counting error, we found that a $(S/N)_{175}$ of $\sim$$50$ is needed to achieve effective measurements using the MIT technique. It is worth mentioning that the effect of other uncertainty sources (e.g., the uncertainty of radiometric calibration) are not included in this study, which should be investigated in the future.  

\begin{acknowledgments}
The authors are supported by the National Key R\&D Program of China No. 2021YFA0718600 and NSFC grant 11825301. This work was carried out using the SWMF/BATSRUS codes developed at the University of Michigan Center for Space Environment Modeling (CSEM) and made available through the NASA Community.
\end{acknowledgments}

\bibliography{ref}{}

\begin{thebibliography}{}
\expandafter\ifx\csname natexlab\endcsname\relax\def\natexlab#1{#1}\fi
\providecommand{\url}[1]{\href{#1}{#1}}
\providecommand{\dodoi}[1]{doi:~\href{http://doi.org/#1}{\nolinkurl{#1}}}
\providecommand{\doeprint}[1]{\href{http://ascl.net/#1}{\nolinkurl{http://ascl.net/#1}}}
\providecommand{\doarXiv}[1]{\href{https://arxiv.org/abs/#1}{\nolinkurl{https://arxiv.org/abs/#1}}}

\bibitem[{{Airapetian} {et~al.}(2017){Airapetian}, {Glocer}, {Khazanov},
  {Loyd}, {France}, {Sojka}, {Danchi}, \& {Liemohn}}]{Airpetian2017}
{Airapetian}, V.~S., {Glocer}, A., {Khazanov}, G.~V., {et~al.} 2017, \apjl,
  836, L3, \dodoi{10.3847/2041-8213/836/1/L3}

\bibitem[{{Alvarado-G{\'o}mez} {et~al.}(2022){Alvarado-G{\'o}mez}, {Drake},
  {Cohen}, {Fraschetti}, {Garraffo}, \& {Poppenh{\"a}ger}}]{Alvarado2022}
{Alvarado-G{\'o}mez}, J.~D., {Drake}, J.~J., {Cohen}, O., {et~al.} 2022,
  Astronomische Nachrichten, 343, e10100, \dodoi{10.1002/asna.20210100}

\bibitem[{{Alvarado-G{\'o}mez} {et~al.}(2018){Alvarado-G{\'o}mez}, {Drake},
  {Cohen}, {Moschou}, \& {Garraffo}}]{Alvarado2018}
{Alvarado-G{\'o}mez}, J.~D., {Drake}, J.~J., {Cohen}, O., {Moschou}, S.~P., \&
  {Garraffo}, C. 2018, \apj, 862, 93, \dodoi{10.3847/1538-4357/aacb7f}

\bibitem[{{Alvarado-G{\'o}mez} {et~al.}(2019){Alvarado-G{\'o}mez}, {Drake},
  {Moschou}, {Garraffo}, {Cohen}, {NASA LWS Focus Science Team: Solar-Stellar
  Connection}, {Yadav}, \& {Fraschetti}}]{Alvarado2019}
{Alvarado-G{\'o}mez}, J.~D., {Drake}, J.~J., {Moschou}, S.~P., {et~al.} 2019,
  \apjl, 884, L13, \dodoi{10.3847/2041-8213/ab44d0}

\bibitem[{{Argiroffi} {et~al.}(2019){Argiroffi}, {Reale}, {Drake},
  {Ciaravella}, {Testa}, {Bonito}, {Miceli}, {Orlando}, \&
  {Peres}}]{Argiroffi2019}
{Argiroffi}, C., {Reale}, F., {Drake}, J.~J., {et~al.} 2019, Nature Astronomy,
  3, 742, \dodoi{10.1038/s41550-019-0781-4}

\bibitem[{{Bondar}(2019)}]{Bondar2019}
{Bondar}, N.~I. 2019, Astronomical and Astrophysical Transactions, 31, 295

\bibitem[{{Brooks} \& {Yardley}(2021)}]{Brooks2021}
{Brooks}, D.~H., \& {Yardley}, S.~L. 2021, Science Advances, 7, eabf0068,
  \dodoi{10.1126/sciadv.abf0068}

\bibitem[{{Brosius} {et~al.}(1998){Brosius}, {Davila}, \&
  {Thomas}}]{Brosius1998}
{Brosius}, J.~W., {Davila}, J.~M., \& {Thomas}, R.~J. 1998, \apjs, 119, 255,
  \dodoi{10.1086/313163}

\bibitem[{{Chen} {et~al.}(2021{\natexlab{a}}){Chen}, {Zhan}, {Youngblood},
  {Wolf}, {Feinstein}, \& {Horton}}]{Chen2021}
{Chen}, H., {Zhan}, Z., {Youngblood}, A., {et~al.} 2021{\natexlab{a}}, Nature
  Astronomy, 5, 298, \dodoi{10.1038/s41550-020-01264-1}

\bibitem[{{Chen} {et~al.}(2021{\natexlab{b}}){Chen}, {Li}, {Tian}, {Chen},
  {Bai}, {Yang}, {Yang}, {Liu}, \& {Deng}}]{Chen2021a}
{Chen}, Y., {Li}, W., {Tian}, H., {et~al.} 2021{\natexlab{b}}, \apj, 920, 116,
  \dodoi{10.3847/1538-4357/ac1792}

\bibitem[{{Chen} {et~al.}(2021{\natexlab{c}}){Chen}, {Liu}, {Tian}, {Bai},
  {Jin}, {Li}, {Yang}, {Yang}, \& {Deng}}]{Chen2021b}
{Chen}, Y., {Liu}, X., {Tian}, H., {et~al.} 2021{\natexlab{c}}, \apjl, 918,
  L13, \dodoi{10.3847/2041-8213/ac1e9a}

\bibitem[{{Cherenkov} {et~al.}(2017){Cherenkov}, {Bisikalo}, {Fossati}, \&
  {M{\"o}stl}}]{Cherenkov2017}
{Cherenkov}, A., {Bisikalo}, D., {Fossati}, L., \& {M{\"o}stl}, C. 2017, \apj,
  846, 31, \dodoi{10.3847/1538-4357/aa82b2}

\bibitem[{{Cohen} {et~al.}(2014){Cohen}, {Drake}, {Glocer}, {Garraffo},
  {Poppenhaeger}, {Bell}, {Ridley}, \& {Gombosi}}]{Cohen2014}
{Cohen}, O., {Drake}, J.~J., {Glocer}, A., {et~al.} 2014, \apj, 790, 57,
  \dodoi{10.1088/0004-637X/790/1/57}

\bibitem[{{Del Zanna} {et~al.}(2021){Del Zanna}, {Dere}, {Young}, \&
  {Landi}}]{DelZanna2021}
{Del Zanna}, G., {Dere}, K.~P., {Young}, P.~R., \& {Landi}, E. 2021, \apj, 909,
  38, \dodoi{10.3847/1538-4357/abd8ce}

\bibitem[{{Del Zanna} \& {Mason}(2018)}]{DelZanna2018}
{Del Zanna}, G., \& {Mason}, H.~E. 2018, Living Reviews in Solar Physics, 15,
  5, \dodoi{10.1007/s41116-018-0015-3}

\bibitem[{{Del Zanna} {et~al.}(2012){Del Zanna}, {Storey}, {Badnell}, \&
  {Mason}}]{DelZanna2012}
{Del Zanna}, G., {Storey}, P.~J., {Badnell}, N.~R., \& {Mason}, H.~E. 2012,
  \aap, 541, A90, \dodoi{10.1051/0004-6361/201118720}

\bibitem[{{Dere} {et~al.}(1997){Dere}, {Landi}, {Mason}, {Monsignori Fossi}, \&
  {Young}}]{Dere1997}
{Dere}, K.~P., {Landi}, E., {Mason}, H.~E., {Monsignori Fossi}, B.~C., \&
  {Young}, P.~R. 1997, \aaps, 125, 149, \dodoi{10.1051/aas:1997368}

\bibitem[{{Donati} {et~al.}(2008){Donati}, {Morin}, {Petit}, {Delfosse},
  {Forveille}, {Auri{\`e}re}, {Cabanac}, {Dintrans}, {Fares}, {Gastine},
  {Jardine}, {Ligni{\`e}res}, {Paletou}, {Ramirez Velez}, \&
  {Th{\'e}ado}}]{Donati2008}
{Donati}, J.~F., {Morin}, J., {Petit}, P., {et~al.} 2008, \mnras, 390, 545,
  \dodoi{10.1111/j.1365-2966.2008.13799.x}

\bibitem[{{Gary} \& {Linsky}(1981)}]{Gary1981}
{Gary}, D.~E., \& {Linsky}, J.~L. 1981, \apj, 250, 284, \dodoi{10.1086/159373}

\bibitem[{{G{\"u}del}(2002)}]{Gudel2002}
{G{\"u}del}, M. 2002, \araa, 40, 217,
  \dodoi{10.1146/annurev.astro.40.060401.093806}

\bibitem[{{Hazra} {et~al.}(2022){Hazra}, {Vidotto}, {Carolan}, {Villarreal
  D'Angelo}, \& {Manchester}}]{Hazra2022}
{Hazra}, G., {Vidotto}, A.~A., {Carolan}, S., {Villarreal D'Angelo}, C., \&
  {Manchester}, W. 2022, \mnras, 509, 5858, \dodoi{10.1093/mnras/stab3271}

\bibitem[{{H{\'e}brard} {et~al.}(2016){H{\'e}brard}, {Donati}, {Delfosse},
  {Morin}, {Moutou}, \& {Boisse}}]{Hebrard2016}
{H{\'e}brard}, {\'E}.~M., {Donati}, J.~F., {Delfosse}, X., {et~al.} 2016,
  \mnras, 461, 1465, \dodoi{10.1093/mnras/stw1346}

\bibitem[{{Hollweg}(1986)}]{Hollweg1986}
{Hollweg}, J.~V. 1986, \jgr, 91, 4111, \dodoi{10.1029/JA091iA04p04111}

\bibitem[{{Jiang} {et~al.}(2022){Jiang}, {Feng}, {Guo}, \& {Hu}}]{Jiang2022}
{Jiang}, C., {Feng}, X., {Guo}, Y., \& {Hu}, Q. 2022, The Innovation, 3,
  100236, \dodoi{10.1016/j.xinn.2022.100236}

\bibitem[{{Jin} {et~al.}(2012){Jin}, {Manchester}, {van der Holst},
  {Gruesbeck}, {Frazin}, {Landi}, {Vasquez}, {Lamy}, {Llebaria}, {Fedorov},
  {Toth}, \& {Gombosi}}]{Jin2012}
{Jin}, M., {Manchester}, W.~B., {van der Holst}, B., {et~al.} 2012, \apj, 745,
  6, \dodoi{10.1088/0004-637X/745/1/6}

\bibitem[{{Johns-Krull} \& {Valenti}(2000)}]{Johns2000}
{Johns-Krull}, C.~M., \& {Valenti}, J.~A. 2000, in Astronomical Society of the
  Pacific Conference Series, Vol. 198, Stellar Clusters and Associations:
  Convection, Rotation, and Dynamos, ed. R.~{Pallavicini}, G.~{Micela}, \&
  S.~{Sciortino}, 371

\bibitem[{{Kochukhov} \& {Lavail}(2017)}]{Kochukhov2017}
{Kochukhov}, O., \& {Lavail}, A. 2017, \apjl, 835, L4,
  \dodoi{10.3847/2041-8213/835/1/L4}

\bibitem[{{Kochukhov} \& {Shulyak}(2019)}]{Kochukhov2019}
{Kochukhov}, O., \& {Shulyak}, D. 2019, \apj, 873, 69,
  \dodoi{10.3847/1538-4357/ab06c5}

\bibitem[{{Landi} {et~al.}(2020){Landi}, {Hutton}, {Brage}, \&
  {Li}}]{Landi2020}
{Landi}, E., {Hutton}, R., {Brage}, T., \& {Li}, W. 2020, \apj, 904, 87,
  \dodoi{10.3847/1538-4357/abbf54}

\bibitem[{{Landi} {et~al.}(2021){Landi}, {Li}, {Brage}, \&
  {Hutton}}]{Landi2021}
{Landi}, E., {Li}, W., {Brage}, T., \& {Hutton}, R. 2021, arXiv e-prints,
  arXiv:2102.06072.
\newblock \doarXiv{2102.06072}

\bibitem[{{Li} {et~al.}(2021){Li}, {Li}, {Wang}, {Brage}, {Hutton}, \&
  {Landi}}]{Li2021}
{Li}, W., {Li}, M., {Wang}, K., {et~al.} 2021, \apj, 913, 135,
  \dodoi{10.3847/1538-4357/abfa97}

\bibitem[{{Li} {et~al.}(2015){Li}, {Grumer}, {Yang}, {Brage}, {Yao}, {Chen},
  {Watanabe}, {J{\"o}nsson}, {Lundstedt}, {Hutton}, \& {Zou}}]{Li2015}
{Li}, W., {Grumer}, J., {Yang}, Y., {et~al.} 2015, \apj, 807, 69,
  \dodoi{10.1088/0004-637X/807/1/69}

\bibitem[{{Li} {et~al.}(2016){Li}, {Yang}, {Tu}, {Xiao}, {Grumer}, {Brage},
  {Watanabe}, {Hutton}, \& {Zou}}]{Li2016}
{Li}, W., {Yang}, Y., {Tu}, B., {et~al.} 2016, \apj, 826, 219,
  \dodoi{10.3847/0004-637X/826/2/219}

\bibitem[{{Linsky}(2019)}]{Linsky2019}
{Linsky}, J. 2019, {Host Stars and their Effects on Exoplanet Atmospheres},
  Vol. 955, \dodoi{10.1007/978-3-030-11452-7}

\bibitem[{{Maehara} {et~al.}(2012){Maehara}, {Shibayama}, {Notsu}, {Notsu},
  {Nagao}, {Kusaba}, {Honda}, {Nogami}, \& {Shibata}}]{Maehara2012}
{Maehara}, H., {Shibayama}, T., {Notsu}, S., {et~al.} 2012, \nat, 485, 478,
  \dodoi{10.1038/nature11063}

\bibitem[{{Morin} {et~al.}(2008){Morin}, {Donati}, {Petit}, {Delfosse},
  {Forveille}, {Albert}, {Auri{\`e}re}, {Cabanac}, {Dintrans}, {Fares},
  {Gastine}, {Jardine}, {Ligni{\`e}res}, {Paletou}, {Ramirez Velez}, \&
  {Th{\'e}ado}}]{Morin2008}
{Morin}, J., {Donati}, J.~F., {Petit}, P., {et~al.} 2008, \mnras, 390, 567,
  \dodoi{10.1111/j.1365-2966.2008.13809.x}

\bibitem[{{Mutel} {et~al.}(1985){Mutel}, {Lestrade}, {Preston}, \&
  {Phillips}}]{Mutel1985}
{Mutel}, R.~L., {Lestrade}, J.~F., {Preston}, R.~A., \& {Phillips}, R.~B. 1985,
  \apj, 289, 262, \dodoi{10.1086/162886}

\bibitem[{{Namekata} {et~al.}(2021){Namekata}, {Maehara}, {Honda}, {Notsu},
  {Okamoto}, {Takahashi}, {Takayama}, {Ohshima}, {Saito}, {Katoh}, {Tozuka},
  {Murata}, {Ogawa}, {Niwano}, {Adachi}, {Oeda}, {Shiraishi}, {Isogai}, {Seki},
  {Ishii}, {Ichimoto}, {Nogami}, \& {Shibata}}]{Namekata2022}
{Namekata}, K., {Maehara}, H., {Honda}, S., {et~al.} 2021, Nature Astronomy, 6,
  241, \dodoi{10.1038/s41550-021-01532-8}

\bibitem[{{Oran} {et~al.}(2013){Oran}, {van der Holst}, {Landi}, {Jin},
  {Sokolov}, \& {Gombosi}}]{Oran2013}
{Oran}, R., {van der Holst}, B., {Landi}, E., {et~al.} 2013, \apj, 778, 176,
  \dodoi{10.1088/0004-637X/778/2/176}

\bibitem[{{Pevtsov} {et~al.}(2003){Pevtsov}, {Fisher}, {Acton}, {Longcope},
  {Johns-Krull}, {Kankelborg}, \& {Metcalf}}]{Pevtsov2003}
{Pevtsov}, A.~A., {Fisher}, G.~H., {Acton}, L.~W., {et~al.} 2003, \apj, 598,
  1387, \dodoi{10.1086/378944}

\bibitem[{{Reiners}(2012)}]{Reiners2012}
{Reiners}, A. 2012, Living Reviews in Solar Physics, 9, 1,
  \dodoi{10.12942/lrsp-2012-1}

\bibitem[{{Semel}(1989)}]{Semel1989}
{Semel}, M. 1989, \aap, 225, 456

\bibitem[{{Si} {et~al.}(2020){Si}, {Brage}, {Li}, {Grumer}, {Li}, \&
  {Hutton}}]{Si2020}
{Si}, R., {Brage}, T., {Li}, W., {et~al.} 2020, \apjl, 898, L34,
  \dodoi{10.3847/2041-8213/aba18c}

\bibitem[{{Sokolov} {et~al.}(2013){Sokolov}, {van der Holst}, {Oran}, {Downs},
  {Roussev}, {Jin}, {Manchester}, {Evans}, \& {Gombosi}}]{Sokolov2013}
{Sokolov}, I.~V., {van der Holst}, B., {Oran}, R., {et~al.} 2013, \apj, 764,
  23, \dodoi{10.1088/0004-637X/764/1/23}

\bibitem[{{Tilley} {et~al.}(2019){Tilley}, {Segura}, {Meadows}, {Hawley}, \&
  {Davenport}}]{Tilley2019}
{Tilley}, M.~A., {Segura}, A., {Meadows}, V., {Hawley}, S., \& {Davenport}, J.
  2019, Astrobiology, 19, 64, \dodoi{10.1089/ast.2017.1794}

\bibitem[{{T{\'o}th} {et~al.}(2012){T{\'o}th}, {van der Holst}, {Sokolov}, {De
  Zeeuw}, {Gombosi}, {Fang}, {Manchester}, {Meng}, {Najib}, {Powell}, {Stout},
  {Glocer}, {Ma}, \& {Opher}}]{SWMF}
{T{\'o}th}, G., {van der Holst}, B., {Sokolov}, I.~V., {et~al.} 2012, Journal
  of Computational Physics, 231, 870, \dodoi{10.1016/j.jcp.2011.02.006}

\bibitem[{{van der Holst} {et~al.}(2014){van der Holst}, {Sokolov}, {Meng},
  {Jin}, {Manchester}, {T{\'o}th}, \& {Gombosi}}]{AWSoM}
{van der Holst}, B., {Sokolov}, I.~V., {Meng}, X., {et~al.} 2014, \apj, 782,
  81, \dodoi{10.1088/0004-637X/782/2/81}

\bibitem[{{Veronig} {et~al.}(2021){Veronig}, {Odert}, {Leitzinger}, {Dissauer},
  {Fleck}, \& {Hudson}}]{Veronig2021}
{Veronig}, A.~M., {Odert}, P., {Leitzinger}, M., {et~al.} 2021, Nature
  Astronomy, 5, 697, \dodoi{10.1038/s41550-021-01345-9}

\bibitem[{{Wang} {et~al.}(2020){Wang}, {J{\"o}nsson}, {Del Zanna}, {Godefroid},
  {Chen}, {Chen}, \& {Yan}}]{Wang2020}
{Wang}, K., {J{\"o}nsson}, P., {Del Zanna}, G., {et~al.} 2020, \apjs, 246, 1,
  \dodoi{10.3847/1538-4365/ab5530}

\bibitem[{{Wilson}(1978)}]{Wilson1978}
{Wilson}, O.~C. 1978, \apj, 226, 379, \dodoi{10.1086/156618}

\bibitem[{{Yang} {et~al.}(2020{\natexlab{a}}){Yang}, {Tian}, {Tomczyk},
  {Morton}, {Bai}, {Samanta}, \& {Chen}}]{Yang2020b}
{Yang}, Z., {Tian}, H., {Tomczyk}, S., {et~al.} 2020{\natexlab{a}}, Science
  China Technological Sciences, 63, 2357, \dodoi{10.1007/s11431-020-1706-9}

\bibitem[{{Yang} {et~al.}(2020{\natexlab{b}}){Yang}, {Bethge}, {Tian},
  {Tomczyk}, {Morton}, {Del Zanna}, {McIntosh}, {Karak}, {Gibson}, {Samanta},
  {He}, {Chen}, \& {Wang}}]{Yang2020a}
{Yang}, Z., {Bethge}, C., {Tian}, H., {et~al.} 2020{\natexlab{b}}, Science,
  369, 694, \dodoi{10.1126/science.abb4462}

\bibitem[{{Zhu} {et~al.}(2022){Zhu}, {Neukirch}, \& {Wiegelmann}}]{Zhu2022}
{Zhu}, X., {Neukirch}, T., \& {Wiegelmann}, T. 2022, Science China
  Technological Sciences, \dodoi{10.1007/s11431-022-2047-8}

\end{thebibliography}
\bibliographystyle{aasjournal}

\end{document}